\documentclass[twocolumn]{aastex631}
\usepackage{fontspec}

\newfontfamily\russianfont{FreeSerif.ttf}

\def\kms{\,{\rm km}\,{\rm s}^{-1}}

\def\feh{\hbox{[Fe/H]}}

\newcommand{\teff}{T_{\rm eff}}
\newcommand{\rv}{{\rm RV}}

\def\vsini{(V \sin{i})}
\def\logg{\log{\rm (g)}}

\begin{document}


\title{J060242.22+332452.7: hierarchical triple system with inner eclipsing contact binary.}

\correspondingauthor{Mikhail Kovalev}

\author[0000-0002-9975-7833]{Mikhail Yu. Kovalev (\russianfont{Михаил~Юрьевич~Ковалёв})}
\email{mikhail.kovalev@ynao.ac.cn}
\affiliation{Yunnan Observatories, China Academy of Sciences, Kunming 650216, China}


\begin{abstract}

I present a discovery of hierarchical triple system J060242.22+332452.7. Inner contact binary system shows clear eclipse timing variation (ETV) due to light time travel effect (LTTE), with LAMOST medium resolution spectra reveals radial velocity (RV) changes of the third star. Joint fit of ETV and RV allows determination of outer orbit with period $P_{\rm out}=1866\pm6$ days and $e_3=0.40\pm0.05$. The masses are $M_{\rm AB,~C}\sin^3{i_3}=2.02\pm0.22,~0.86\pm0.08~M_\odot$. This system is expected to have an astrometric orbit in the next data release of Gaia. 

\end{abstract}



\section{Introduction} \label{sec:intro}

Recently, an interesting compact hierarchical triple (CHT) J04+25 was discovered by \cite{ttau}, using spectra from LAMOST medium resolution survey (MRS \cite{mrs}). It contained surprisingly bright third star ($\sim70\%$ of total light), which outshines inner eclipsing contact binary. Further analysis of the LAMOST-MRS spectra by \cite{v885per} revealed another interesting system, where third spectral component was found to orbit known eclipsing contact binary V 885 Per every $\sim21.8$ days, which made it most compact outer orbit currently known in triple systems. Such amazing discoveries have proven, that LAMOST-MRS spectra are very useful in the search of third stars around contact binaries. In this Note, I report LAMOST-MRS based discovery of another hierarchical triple system: J060242.22+332452.7, which is quite common in comparison with two previous systems, although it has longest orbital period among systems discovered by the author.

\section{DATA and Methods}
\subsection{spectra}
All available spectra of J060242.22+332452.7 were downloaded from \url{lamost.org/dr11}.
Similarly to previous studies \citep{ttau,v885per}, I used only individual spectra, taken with exposure times of 20 minutes. Following same ``Matryoshka" approach, each spectrum was fitted by the binary spectral model in two iterations: at first brighter spectral component is found in first iteration and subtracted from the spectrum, which was fitted again in the second iteration, to get second and third spectral components. Thus, radial velocities and spectral parameters were derived for all three stars of the triple system for each spectrum. Time averages of spectral parameters are following:
\begin{itemize}
    \item ${\teff}_{A,B,C}=6273\pm60,6175\pm97,  6023\pm317$ K,
    \item $\logg_{A,B,C} =4.15\pm0.24 ,3.71\pm0.55,  3.79\pm0.81$ dex,
    \item $\feh_{AB,C}=-0.16\pm0.08, -0.13\pm0.14$ dex,
    \item $\vsini_{A,B,C}=153\pm23 ,182\pm52,21\pm47\,\kms$
\end{itemize}
  
where indexes $A,B$ corresponds to primary and secondary in the inner contact binary, index $C$ is a third star. $\feh_{AB}$ is a common metallicity for a contact system.
It is clear that components in the inner compact binary have much faster rotation, than the third star.

The upper right panel of the Figure~\ref{fig:spexampl} shows $\rv$ for all components versus phase of the inner eclipsing system computed using ephemeris from \cite{asas12k}:
\begin{equation}
   t_{min}(\rm HJD)=2458112.3175+0.48378\, E, 
\label{eq:efimer} 
\end{equation}
    
Based on this plot we can see that components of contact binary shows typical sinusoidal $\rv$ curves, consistent with circular orbits, typical for contact binaries, with third star showing noticeable $\rv$ changes relative to the common center of mass. Mass ratio in inner contact system is $q_{\rm in}\sim0.3$. The sum of $\rv$ semiamplitudes is $K_A+K_B\sim320~\kms$.

\subsection{photometry}
\begin{figure*}
    \includegraphics[width=0.48\linewidth]{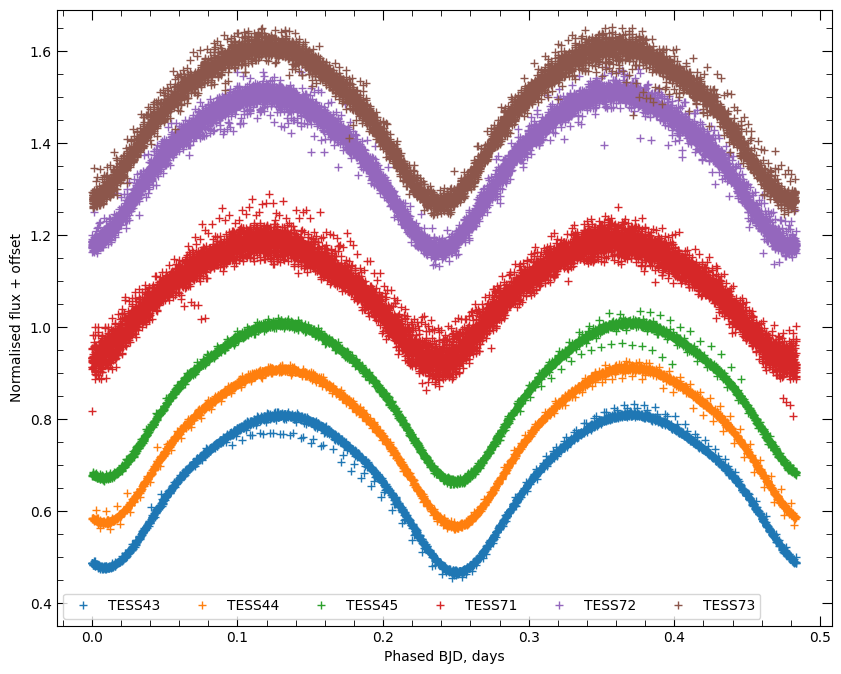}
    \includegraphics[width=0.52\linewidth]{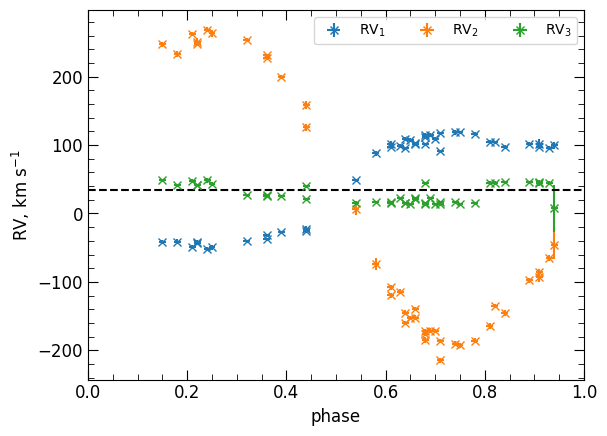}\\
    \includegraphics[width=0.99\linewidth]{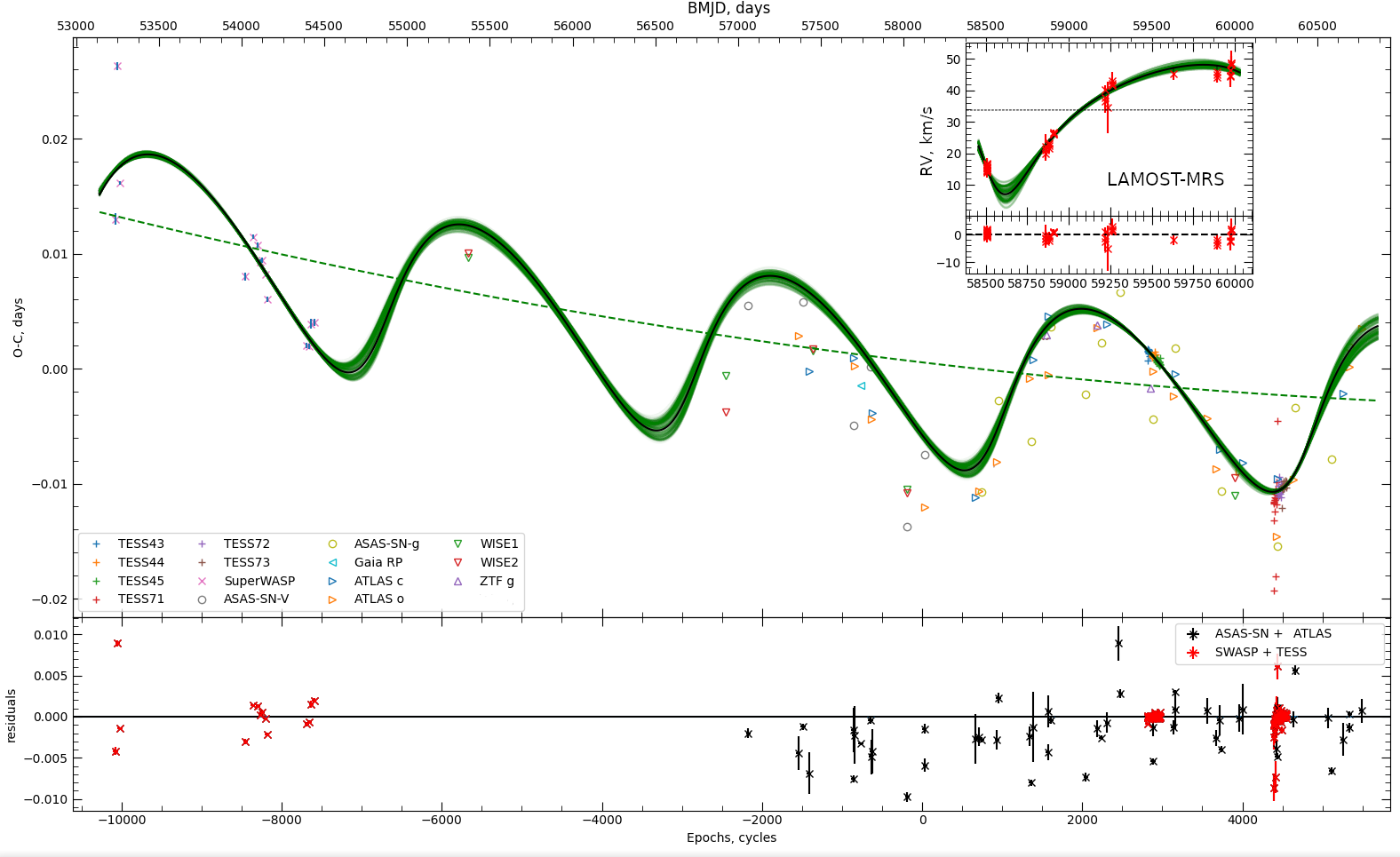}
    \caption{ TESS light curves for six sectors, showing obvious LTTE shift (top left panel). RV from LAMOST-MRS for all three components (top right panel). Joint fit of LTTE shifts and $\rv_3$ (inline plot) is shown in the bottom panel.} 
    \label{fig:spexampl}
\end{figure*}

I downloaded photometry from public surveys including SuperWASP \citep{superwasp}, ZTF \citep{ztf_main}, Gaia DR3 \citep{gaia3}, ASAS-SN \citep{asassnv}, ATLAS \citep{atlas1}, WISE \citep{wise} and TESS \citep{tess}. Later one observed this target during sectors 43,44,45 and 71,72,73, each lasting for one month. For all sectors I used Quick Look Pipeline simple aperture photometry \citep{tess1,tess2}, except for sector 71, where TESS-Gaia Light Curve pipeline \citep{tglc} was used due to significantly higher sky background. Upper left panel of Figure~\ref{fig:spexampl} shows these light curves (LCs), normalized and phased using ephemeris from Equation~\ref{eq:efimer}. Eclipses have nearly identical depth, also it is evident, that times of minima show significant changes with time, which can be explained by the LTTE, caused by orbital motion of contact binary barycenter \citep{ltte}.
\par
To find eclipse time variation, I employed template fitting of LCs, similarly to \cite{ttau}. Wilson–Devinney program {\sc W-D} \citep{wd71,wilson79,wilson14} with {\sc PYWD2015} \citep{pywd} user interface were used to make a ``toy" model of LCs in TESS, ASAS-SN and SuperWASP bands. All LCs show noticeable third light contribution $L_3\sim5-11\%$. Taking these LCs as templates, I found timeshifts, required to match observations. These timeshifts are essentially the same as O-C values, commonly used to characterize period change and LTTE in eclipsing systems.
\par
The O-C curve and third star $\rv_3$ were modeled using composite model based on Keplerian orbit from PyAstronomy\citep{pyasl}, which allows the computation of position and velocity along the line of sight for a two-body problem. Model also includes parabolic terms for ETV, which take into account possible period change and corrections to linear ephemeris. To find solution, I used $\chi^2$ minimisation with ETV points selected only from TESS and SuperWASP data, because they were measured in relatively short time intervals (2 and 20 days for TESS and SuperWASP respectively).
When best solution was found, I run Markov Chain Monte Carlo (MCMC) sampling with {\sc EMCEE} \citep{mcmc} using 48 walkers and 27500 iterations. The resulting models for O-C and $\rv_3$ curves are shown in lower panel of Figure~\ref{fig:spexampl} as solid black lines, where I also show O-C values from ZTF, WISE, ASAS-SN, Gaia and ATLAS just for reference. A hundred Keplerian orbits, randomly selected from sampled solutions, are also shown as green lines. Dashed green line shows parabolic trend. 
Final solution (computed as median and standard deviation of the sample) is following:
\begin{itemize}
    \item $\Delta t_0=-0.0026\pm0.0005$ day, 
    \item $\Delta P =-0.74\pm0.09\cdot 10^{-6}$ day cycle$^{-1}$,
    \item $\beta=\frac{dP}{dt} = 5.4\pm1.3\cdot 10^{-11}$ day cycle$^{-2}$,
    \item $\mathcal{A}=   0.0080\pm0.0004$ day, 
    \item $P_3=  1866\pm6$ day,  
    \item $t_{p3}=  58522.03\pm26.83$ day (BMJD),  
    \item $e_3=  0.40\pm0.05$,
    \item ${\omega_3}  = -31.2\pm5.3^\circ$,
    \item $q_3=\frac{M_A+M_B}{M_C}=2.34\pm0.07$,
    \item $\gamma_3=34.63\pm0.37\,\kms$,
\end{itemize}

where $\Delta t_0,~\Delta P$ are corrections for linear ephemeris, $\beta$ is a long-term change of the period, $\mathcal{A}\equiv a/c$ is an amplitude of LTTE (semimajor axis devided by the speed of light), $P_3, t_{p3}, e_3, {\omega_3}, q_3, \gamma_3$ are period, periastron passage time, eccentricity, periastron argument, mass ratio and systemic velocity for the outer orbit.   


\section{Discussion}

Outer orbit parameters allow to make estimation of the system's minimal mass ($M_{\rm tot}=M_A+M_B+M_C$) using Kepler's third law:
\begin{equation}
        M_{\rm tot}\sin^3{i_3}=\frac{\left(c~\mathcal{A}~(1+q_3)\right)^3}{{GM_{\odot}}} \left[\frac{2 \pi}{P_3}\right]^2=2.88\pm0.30~M_\odot ,
    \label{eq:kepler3}
\end{equation}%

where $c$ - is a speed of light,  $GM_\odot=1.32712440041\cdot 10^{20}\, {\rm m^3\,s^{-2}}$ - is the Solar mass parameter\footnote{\url{https://iau-a3.gitlab.io/NSFA/NSFA_cbe.html\#GMS2012}}. Thus, the contact system and the third star have masses $M_{ AB,~C}\sin^3{i_3}=2.02\pm0.22,~0.86\pm0.08~M_\odot$. 
Inner system's total mass can be roughly estimated based on the sum of $\rv$ semiamplitudes, period, inclination angle $i_{AB}=69^\circ$ \citep{asas12k} and  Kepler's third law:

\begin{equation}
    M_{AB}=M_A+M_B= \frac{(K_A+K_B)^3}{\sin^3 i_{AB}}\frac{P_{AB}}{2\pi\,GM_\odot} \sim 1.97~M_\odot
\end{equation}

Combining these two values of $M_{AB}$ we can expect that outer orbit's inclination angle is relatively $i_3\sim80-90^\circ$, so the outer orbit is nearly perpendicular to the sky-plane.  
\par
Projected semimajor axes for the contact system and the third star are: $a_{AB,~C} \sin{i_3}=297\pm14,~694\pm37~R_\odot$. The maximal angular separation between the contact system and the third  star is around $\sim9$ mas, if we use the Gaia DR3 distance $d=671$ pc, which is too small to be resolved with currently available technology. However, systems's photocenter should move relative to the center of mass with amplitude $\sim2$ mas, slightly oscillating due to the variability of the inner eclipsing system. Gaia DR3 Non-single star orbital catalog \citep{gaia3nss} doesn't provide astrometric orbit, similarly to case of J04+25 \citep{ttau}. Renormalised unit weight error RUWE=$1.523$ is relatively small, probably because Gaia DR3 is based on observations with a time base covering only a half of orbital period for third star. Hopefully, upcoming Gaia DR4 will provide an orbital solution for J060242.22+332452.7, so one can get an independent estimation of inclination for the outer orbit.
\par
Eccentricity of the outer orbit have large uncertainty, mainly because LAMOST-MRS spectra have a poor coverage near the periastron passage. 
Future spectroscopic observations are required to be scheduled for this important part of the orbit, which can significantly improve eccentricity estimation. Next periastron passage will happened at MJD$\sim62254\pm32$ d (around 2029-04-28, within a month) and should be easily observable by LAMOST.

\section{Conclusions}

I present a discovery of another hierarchical triple system, containing eclipsing contact binary. LAMOST-MRS spectra allow to measure the third star's $\rv$ curve, which is consistent with the ``light" orbit obtained using ETV for inner stellar pair. Derived orbit has $P_{\rm out}=1866\pm6$ days and moderate $e=0.40\pm0.05$. Based on comparison of mass estimates for the inner system outer orbit should be nearly perpendicular to the sky-plane. Gaia DR3 didn't provide astrometric orbit for this system, however I expect that next Gaia DR4 will have it, since it is based on observations covering one full outer orbit. Future spectroscopic observation are required to constrain the eccentricity of the outer orbit.

\begin{acknowledgments}

This work made use of the data from LAMOST (Large Sky Area Multi-Object Fiber Spectroscopic Telescope, also known as the Guoshoujing Telescope \cite{2012RAA....12.1197C, 2012RAA....12..723Z}) (\url{https://cstr.cn/31118.02.LAMOST}). LAMOST is a Chinese national mega-science facility, operated by National Astronomical Observatories, Chinese Academy of Sciences.
\par
This research has made use of NASA's Astrophysics Data System, the SIMBAD data base, and the VizieR catalogue access tool, operated at CDS, Strasbourg, France. It also made use of TOPCAT, an interactive graphical viewer and editor for tabular data \citep[][]{topcat}.  
Funding for the TESS mission is provided by NASA's Science Mission directorate. This paper includes data collected by the TESS mission, which is publicly available from the Mikulski Archive for Space Telescopes (MAST).
\par
This work has made use of data from the European Space Agency (ESA) mission {\it Gaia} (\url{https://www.cosmos.esa.int/gaia}), processed by the {\it Gaia}
Data Processing and Analysis Consortium (DPAC,
\url{https://www.cosmos.esa.int/web/gaia/dpac/consortium}). Funding for the DPAC
has been provided by national institutions, in particular the institutions
participating in the {\it Gaia} Multilateral Agreement.

\end{acknowledgments}

\bibliographystyle{aasjournal}


\end{document}